\documentclass[journal]{IEEEtran}
\usepackage{mathtools}
\usepackage{cite} 
\usepackage{booktabs}
\usepackage{algorithm}
\usepackage{algpseudocode}
\usepackage{amssymb} 
\usepackage{siunitx} 
\usepackage{graphicx} 
\usepackage{amsmath}  

\begin{document}

\title{Optimizing Resource Allocation for QoS and Stability in Dynamic VLC-NOMA Networks via MARL}

\author{
    Aubida A. Al-Hameed, 
    Safwan Hafeedh Younus, 
    Mohamad A. Ahmed, 
    and Abdullah Baz 
    \thanks{Aubida A. Al-Hameed, Safwan Hafeedh Younus, and Mohamad A. Ahmed are with the Department of Communication Engineering, College of Electronics Engineering, Ninevah University, Mosul, Iraq }
    \thanks{Abdullah Baz is with the Department of Computer and Network Engineering, College of Computing, Umm Al-Qura University, Makkah, Saudi Arabia}%
    \thanks{Corresponding author: Aubida A. Al-Hameed (e-mail:aubida.alhameed@uoninevah.edu.iq)}.
}

\maketitle

\begin{abstract}
Visible Light Communication (VLC) combined with Non-Orthogonal Multiple Access (NOMA) offers a promising solution for dense indoor wireless networks. Yet, managing resources effectively is challenged by VLC network dynamic conditions involving user mobility and light dimming. In addition to satisfying Quality of Service (QoS) and network stability requirements. Traditional resource allocation methods and simpler RL approaches struggle to jointly optimize QoS and stability under the dynamic conditions of mobile VLC-NOMA networks. This paper presents MARL frameworks tailored to perform complex joint optimization of resource allocation (NOMA power, user scheduling) and network stability (interference, handovers), considering heterogeneous QoS, user mobility, and dimming in VLC-NOMA systems. Our MARL frameworks capture dynamic channel conditions and diverse user QoS , enabling effective joint optimization. In these frameworks, VLC access points (APs) act as intelligent agents, learning to allocate power and schedule users to satisfy diverse requirements while maintaining network stability by managing interference and minimizing disruptive handovers. We conduct a comparative analysis of two key MARL paradigms: 1) Centralized Training with Decentralized Execution (CTDE) and 2) Centralized Training with Centralized Execution (CTCE). Comprehensive simulations validate the effectiveness of both tailored MARL frameworks and demonstrate an ability to handle complex optimization. The results show key trade-offs, as the CTDE approach achieved approximately 16\% higher for High priority (HP) user QoS satisfaction, while the CTCE approach yielded nearly 7 dB higher average SINR and 12\% lower ping-pong handover ratio, offering valuable insights into the performance differences between these paradigms in complex VLC-NOMA network scenarios.
\end{abstract}

\section{Introduction}
\label{sec:introduction}

\IEEEPARstart{T}{he} relentless demand for increased data rates and broad connectivity, especially in indoor environments, is propelling research beyond traditional Radio Frequency (RF) wireless technologies. Visible Light Communication (VLC) has arisen as an innovative complementary technology, utilizing current lighting infrastructure to deliver high-bandwidth, license-free, and intrinsically secure wireless connections \cite{Haas2016LiFi} \cite{8576503}. By employing Light Emitting Diodes (LEDs) for illumination as well as data transmission, Visible Light Communication (VLC) provides a substantial unregulated spectrum and significant spatial reuse potential, making it appropriate for densely populated indoor environments \cite{Ghassemlooy2017OWCBook}. Non-Orthogonal Multiple Access (NOMA) has been recognized as a pivotal technology for augmenting spectral efficiency in the VLC systems \cite{Ding2017NOMASurvey} \cite{Younus2019}. NOMA enables multiple users to simultaneously utilize identical time-frequency resources through power-domain multiplexing, considerably enhancing user capacity relative to conventional Orthogonal Multiple Access (OMA) techniques, and is particularly compatible with the characteristics of VLC channels \cite{Marshoud2016NOMAVLC}.

Nonetheless, achieving the entire capacity of VLC-NOMA in dynamic indoor environments poses major challenges. These environments are defined by multiple light sources and users in close proximity, resulting in extensive interference patterns. User mobility and light attenuation lead to channel fluctuations, requiring effective handover strategies between VLC access points to ensure uninterrupted connectivity \cite{Wang2019VLCMobilitySurvey}. In addition, within an indoor network, users often require diverse Quality of Service (QoS) requirements, with applications such as video conferencing and virtual reality demanding elevated data rates and low latency. Alongside traditional data users requiring a minimum assured throughput, this addresses the many requirements expected in forthcoming 6G applications \cite{Chen20206GVision}. Simultaneously guaranteeing higher Quality of Service (QoS) for priority users, managing inter-cell interference (ICI), and facilitating seamless handovers constitutes a highly intricate and multi-faceted resource allocation challenge \cite{Yang20196GVision}. The inherent dependency of these aims, such as maximizing throughput, can cause interference, consequently complicating optimization significantly.

Traditional resource allocation approaches to VLC-NOMA, typically reliant on static rules \cite{Zhang2017VLCUserGrouping}. Also, the mathematical optimization designed for simplified scenarios without with the high dimensions and unpredictability inherent to dynamic VLC-NOMA networks \cite{Akyildiz2005NextGenSurvey}. Optimization methods may become computationally unfeasible, while heuristics frequently yield sub optimal results, struggling to adapt to real-time fluctuations in network conditions, user requirements, or fluctuations in channel quality due to mobility and dimming \cite{Khisti2011ResourceAllocChallenges}. Single-agent Reinforcement Learning (RL) approaches have shown potential in adapting to dynamic conditions in wireless networks \cite{Luong2019DRLSurvey} \cite{Qazzaz2023} \cite{Qazzaz2024}. Early study utilizing reinforcement learning for VLC-NOMA power allocation underlines its potential, however often focuses on relatively basic reinforcement learning models \cite{Guo2020QlearningVLC}. Consequently, a distinct need emerges for an intelligent and coordinated resource allocation mechanism that can develop near-optimal strategies for these sophisticated network requirements.

In this paper, we present a Multi-Agent Reinforcement Learning (MARL) framework to overcome these limitations in resource allocation for dynamic VLC-NOMA networks. MARL is effective at handling challenges involving numerous interacting decision-makers within a such an environment \cite{Busoniu2008MARLSurvey}. In addition, in\cite{Shafi2022MARLUAVVLC}, the MARL has proven effective in complicated resource management tasks, including those in VLC network setups. In our proposed framework, individual VLC access points act as autonomous agents, acquiring cooperative strategies to enhance resource allocation (NOMA power coefficients, user scheduling decisions) based on local observations during training. Our aim is to empower agents to develop policies that collectively enhance network performance across various dimensions: 1) fulfilling diverse QoS needs, 2) maintaining network stability, 3) effectively managing interference, and 4) assuring handover stability. We examine and compare two key MARL paradigms: Centralized Training with Decentralized Execution (CTDE) and Centralized Training with Centralized Execution (CTCE). These paradigms present unique trade-offs between coordination capabilities and execution complexity \cite{Sadeghi2023MARLComNetSurvey}\cite{Ma2024MATD3AoI}. We develop tailored state representations, action spaces, and multi-objective reward functions that explicitly integrate key components of the VLC-NOMA dynamic environment, covering user mobility, light dimming, QoS classes, interference levels, and handover.
This paper's main contributions are summarized as follows:
\begin{itemize}
    \item We propose and assess tailored MARL frameworks, specifically Centralized Training with Decentralized Execution (CTDE) employing Multi-Agent Proximal Policy Optimization (MAPPO) and Centralized Training with Centralized Execution (CTCE) utilizing Centralized Proximal Policy Optimization (PPO), for the tangled task of simultaneously optimizing Quality of Service (QoS) maintaining and network stability, which includes resource allocation, interference management, and handover stability, within dynamic indoor VLC-NOMA networks characterized by user mobility and light dimming.
    \item We developed tailored MARL components, which include state representations that integrate local and neighboring information, composite action spaces that facilitate simultaneous management of NOMA power allocation, users' scheduling and handover triggers. In addition to design a multi-objective reward function that specifically targets those distinct challenges and dynamics of mobile VLC-NOMA systems with diverse user requirements.
    \item We establish a systematic method employing Bayesian Optimization to determine the critical convergence between possibly conflicting QoS and stability objectives within the MARL reward function, thereby providing a methodical tuning procedure for the multi-objective problem.
    \item We illustrated through thorough simulations the ability of the proposed MARL frameworks in attaining enhanced performance in QoS satisfaction, network throughput, fairness, and handover stability compared to a traditional baseline resource allocation strategy.
\end{itemize}

The subsequent sections of this paper are structured as follows. Section \ref{sec:system_model} outlines the system model for the dynamic mobile VLC-NOMA network. Section \ref{sec:marl_framework} defines the proposed MARL framework and the particulars of the CTDE and CTCE implementations. Section \ref{sec:simulation_setup} defines the simulation configuration and settings. Section \ref{sec:results} discusses and analyzes the simulation outcomes. Ultimately, Section \ref{sec:conclusion} concludes the findings and proposes paths for future research.

\section{System Model}
\label{sec:system_model}

\subsection{Network Architecture and Layout}
\label{subsec:network_layout}

The VLC simulation environment is represented as a singular rectangular room with dimensions $L \times W \times H$ meters, displaying an ordinary indoor space such as an office or conference room.

This room has $N_{\mathrm{AP}}$ VLC Access Points (APs), given by the set $\mathcal{A} = \{1, 2, ..., N_{\mathrm{AP}}\}$. The VLC APs are assumed to be mounted to the ceiling at height $H$. The VLC APs are organized in a systematic grid layout to ensure consistent illumination and communication coverage. The 3D coordinates of the $a$-th access point are given as $\mathbf{p}_{a}^{\mathrm{AP}} = (x_a^{\mathrm{AP}}, y_a^{\mathrm{AP}}, H)$, for all $a \in \mathcal{A}$. Each AP is equipped with an LED array that facilitates simultaneous illumination and data transfer using intensity modulation. These VLC APs act as learning agents within our proposed MARL frameworks.

The network accommodates $N_{\mathrm{UE}}$ mobile users, given by the set $\mathcal{U} = \{1, 2, ..., N_{\mathrm{UE}}\}$. It is assumed that users navigate through the room only on a horizontal plane at a standard receiving height $H_{\mathrm{UE}}$ (e.g., desk level, $0 < H_{\mathrm{UE}} < H$). The three-dimensional coordinates of user $u$ at a particular time $t$ are expressed as $\mathbf{p}_{u}^{\mathrm{UE}}(t) = (x_u^{\mathrm{UE}}(t), y_u^{\mathrm{UE}}(t), H_{\mathrm{UE}})$, for all $u \in \mathcal{U}$. Each user has a vertically oriented photodiode (PD) receiver to capture the downlink VLC signals. The characteristics of user mobility are given in Section \ref{subsec:mobility_model}.

\subsection{VLC Channel Model}
\label{subsec:channel_model}

The communication link between each VLC AP $a \in \mathcal{A}$ and user $u \in \mathcal{U}$ is established via the VLC channel. The received optical power at user $u$ from VLC AP $a$ depends on the channel's DC gain $H_{a,u}(t)$, which is time-varying due to user mobility. We consider both Line-of-Sight (LoS) and Non-LoS (NLoS) components.

The LoS channel DC gain $H_{a,u}^{\mathrm{LoS}}(t)$ between VLC AP $a$ located at $\mathbf{p}_{a}^{\mathrm{AP}}$ and user $u$ located at $\mathbf{p}_{u}^{\mathrm{UE}}(t)$ can be modeled using the generalized Lambertian emission pattern \cite{Komine2004VLCChannel} \cite{Al-Hameed2019}. It is given by:
\begin{equation}
\label{eq:h_los}
H_{a,u}^{\mathrm{LoS}}(t) =
\begin{cases}
\begin{aligned}
&\frac{(m+1) A_{\mathrm{PD}}}{2 \pi d_{a,u}^2(t)} \\
&\quad \times \cos^m(\phi_{a,u}(t)) \\
&\quad \times T_s(\psi_{a,u}(t)) \\
&\quad \times g(\psi_{a,u}(t)) \cos(\psi_{a,u}(t)),
\end{aligned}
& 0 \leq \psi_{a,u}(t) \leq \Psi_c \\
0, & \psi_{a,u}(t) > \Psi_c
\end{cases}
\end{equation}

where:
\begin{itemize}
    \item $d_{a,u}(t) = \| \mathbf{p}_{u}^{\mathrm{UE}}(t) - \mathbf{p}_{a}^{\mathrm{AP}} \|_2$ is the Euclidean distance between AP $a$ and user $u$ at time $t$.
    \item $m = -\ln(2) / \ln(\cos(\Phi_{1/2}))$ is the Lambertian order related to the transmitter's semi-angle at half power $\Phi_{1/2}$.
    \item $A_{\mathrm{PD}}$ is the physical detection area of the user's photodiode (PD).
    \item $\phi_{a,u}(t)$ is the angle of irradiance from VLC AP $a$ relative to its perpendicular axis (normal vector).
    \item $\psi_{a,u}(t)$ is the angle of incidence at user $u$'s receiver relative to its perpendicular axis (normal vector).
    \item $T_s(\psi_{a,u}(t))$ is the gain of the optical filter at the receiver.
    \item $g(\psi_{a,u}(t))$ is the gain of the optical concentrator at the receiver.
    \item $\Psi_c$ is the receiver's Field of View (FOV) angle, defining the acceptance angle of the PD.
\end{itemize}
Both angles $\phi_{a,u}(t)$ and $\psi_{a,u}(t)$ are functions of the VLC AP and UE positions and their orientations.

The NLoS components arise primarily from reflections off surfaces within the room. In this work, we model the NLoS contribution by considering only the first-order reflections, denoted by $H_{a,u}^{(1)}(t)$. This component is calculated by integrating the power received from all reflecting surface elements $dA_{\mathrm{ref}}$ within the environment. Assuming Lambertian reflection with reflectivity $\rho$ from each surface element, the first-order reflection gain is \cite{Kahn1997WirelessIRComm, Komine2004VLCChannel}:

\begin{align}
H_{a,u}^{(1)}(t) &= \int_{\text{SURFACES}}
\frac{(m+1)\, A_{\mathrm{PD}}\, \rho}{2 \pi^2\, d_{1}^2(t)\, d_{2}^2(t)}
\cos^m(\phi_{1}(t)) \notag \\ &\quad \times \cos(\psi_{1}(t))
  \cos(\phi_{2}(t))\, \cos(\psi_{2}(t))\, T_S(\psi_{2}(t))  \notag \\ &\quad \times g(\psi_{2}(t)) \;  \mathbb{I}(\psi_2(t) \le \Psi_c) \; dA_{\mathrm{REF}}
\label{EQ:H_NLOS1_INTEGRAL}
\end{align}

where $d_{1}(t)$ and $d_{2}(t)$ are the distances from the VLC AP to $dA_{\mathrm{ref}}$ and from $dA_{\mathrm{ref}}$ to the user, respectively; $\phi_{1}(t)$ and $\psi_{1}(t)$ are the angles of irradiance and incidence for the path from VLC AP to $dA_{\mathrm{ref}}$; $\phi_{2}(t)$ and $\psi_{2}(t)$ are the angles for the path from $dA_{\mathrm{ref}}$ to the user; $\mathbb{I}(\cdot)$ is the indicator function; and the integral is over all reflecting surfaces. While higher-order reflections also contribute to the total received power, however, modelling the first-order reflections captures the most significant portion of the NLoS power while managing computational complexity \cite{Barry1993Multipath, Lopez2011NLOSSurvey}.

The total channel DC gain used in this paper is given as:
\begin{equation}
\label{eq:h_total_detailed_nlos}
H_{a,u}(t) = H_{a,u}^{\mathrm{LoS}}(t) + H_{a,u}^{(1)}(t)
\end{equation}

\subsection{NOMA Transmission and Interference Model}
\label{subsec:noma_model}

In this work, we employ power-domain Non-Orthogonal Multiple Access (NOMA) in the downlink to enhance spectral efficiency \cite{Ding2017NOMASurvey, Marshoud2016NOMAVLC}. Each VLC AP $a \in \mathcal{A}$ serves a dynamically selected user set $\mathcal{U}_a(t) \subseteq \mathcal{U}$ by superimposing their signals $x_{a,u}(t)$ using allocated electrical powers $P_{a,u}(t)$. The total allocated electrical power per VLC AP is constrained by $P_{a,\mathrm{max}}^{\mathrm{elec}}$:
\begin{equation}
\label{eq:power_constraint}
\sum_{u \in \mathcal{U}_a(t)} P_{a,u}(t) \leq P_{a,\mathrm{max}}^{\mathrm{elec}}, \quad \forall a \in \mathcal{A}.
\end{equation}
Intensity Modulation/Direct Detection (IM/DD) is assumed for optical transmission \cite{Kahn1997WirelessIRComm}.

In this work, we assumed that the users' receiver performs a perfect Successive Interference Cancellation (SIC) to decode their messages \cite{Dai2015NOMASurvey}. Users in $\mathcal{U}_a(t)$ are decoded sequentially, typically based on their channel power gains $G_{a,u}(t) = (R H_{a,u}(t))^2$, where $R$ is the PD responsivity and $H_{a,u}(t)$ is the total channel gain from Eq. \eqref{eq:h_total_detailed_nlos}. Assuming perfect SIC, user $u_k$ (where users are indexed $k=1...K_a(t)$ ) treats signals intended for users $j>k$ (users with weaker channels) as noise and cancels the signals from users $j<k$ (users with stronger channels). The resulting Signal-to-Interference-plus-Noise Ratio (SINR) for user $u_k$ served by VLC AP $a$ is:
\begin{equation}
\label{eq:sinr}
\resizebox{0.9\hsize}{!}{$
\mathrm{SINR}_{a,u_k}(t) = \frac{G_{a,u_k}(t) P_{a,u_k}(t)}{\sum_{j \neq k}^{K_a(t)} G_{a,u_k}(t) P_{a,u_j}(t) + I_{\mathrm{ICI}, u_k}(t) + \sigma_{\mathrm{noise}}^2}
$}
\end{equation}
where $\sigma_{\mathrm{noise}}^2$ is the noise variance, and $I_{\mathrm{ICI}, u_k}(t)$ is the Inter-Cell Interference (ICI) from neighboring VLC APs, which has a significant impact in dense VLC-NOMA network deployments \cite{Cui2017ICICVLC} \cite{Liu2017NOMAPAFairness}. The ICI is given as:
\begin{equation}
\label{eq:ici}
I_{\mathrm{ICI}, u_k}(t) = \sum_{a' \in \mathcal{A}, a' \neq a} \sum_{u' \in \mathcal{U}_{a'}(t)} G_{a',u_k}(t) P_{a',u'}(t).
\end{equation}
Here, $G_{a',u_k}(t) = (R H_{a',u_k}(t))^2$ is the channel power gain from interfering/neighboring VLC AP $a'$ to user $u_k$.

The achievable data rate for user $u_k$ is calculated using the Shannon capacity formula over bandwidth $B$:
\begin{equation}
\label{eq:rate}
\mathrm{Rate}_{a,u_k}(t) = B \log_2(1 + \mathrm{SINR}_{a,u_k}(t)).
\end{equation}

Crucially, the ICI term \eqref{eq:ici} couples the decisions (user selection $\mathcal{U}_{a'}(t)$ and power allocation $P_{a',u'}(t)$) made by neighboring VLC APs $a'$. This coupling, combined with the time-varying channel gains $G_{a,u}(t)$ due to mobility and the need to satisfy diverse QoS requirements through careful power allocation. The MARL frameworks proposed in Section \ref{sec:marl_framework} are designed precisely to enable the distributed VLC AP agents to learn effective policies for user selection $\mathcal{U}_a(t)$ and power allocation $P_{a,u}(t)$ that navigate these challenges to jointly optimize QoS and network stability objectives.

\subsection{User Mobility Model}
\label{subsec:mobility_model}

In this study, we deploy the Random Direction (RD) model. The RD model is selected for multiple reasons relevant to our examination of dynamic indoor VLC-NOMA networks.
 1) It allows continuous user mobility, essential for modelling the dynamic fluctuations in channel gains and SINR characteristics of the indoor VLC environment.
 2) In contrast to the Random Waypoint model, RD generally yields a more uniform spatial distribution of users over time, hence preventing unrealistic clustering in the middle of the room \cite{Camp2002MobilitySurvey} \cite{Bettstetter2004RWPLimitations}.
 In the RD model, each user $u \in \mathcal{U}$ initiates at a random location $\mathbf{p}_{u}^{\mathrm{UE}}(0)$ within the room at a height of $H_{\mathrm{UE}}$. The user uniformly selects a starting direction $\theta_u$ from the interval $[0, 2\pi)$ and a speed $v_u$ uniformly from the range $[v_{\mathrm{min}}, v_{\mathrm{max}}]$, which denotes normal pedestrian velocities. The user thereafter proceeds in a linear trajectory with the velocity vector $\mathbf{v}_u = (v_u \cos \theta_u, v_u \sin \theta_u, 0)$ until encountering one of the room's boundaries.

 Once the user approach one of the room's boundaries, the user's trajectory changes and a new velocity and direction are selected. For the sake of simplicity, we assume there is no stop period at the boundaries. The user continues in this procedure for the entirety of the simulation period.

 User locations are updated at discrete time periods $\Delta t$. The position at the subsequent time step is determined as:
\begin{equation}
\label{eq:position_update}
\mathbf{p}_{u}^{\mathrm{UE}}(t + \Delta t) = \mathbf{p}_{u}^{\mathrm{UE}}(t) + \mathbf{v}_u(t) \cdot \Delta t
\end{equation}
where $\mathbf{v}_u(t)$ is the user's velocity vector during the interval $[t, t+\Delta t)$, potentially changing at room's boundaries.

\subsection{Network Quality of Service (QoS)}
\label{subsec:qos_model}

This section outlines the service requirements of users within the simulated VLC-NOMA network which is reflecting a realistic indoor application demand. Achieving these diverse QoS requirements while maintaining network stability forms the primary challenge addressed in this study through the proposed MARL-based resource allocation strategy.

We group users into two distinct groups depending on their application requirements: \begin{itemize}
\item{High-Priority (HP) Users:} Representing users using high-bandwidth/data rate services, for instance, real-time video conferencing or interactive gaming. The group of HP users is given as $\mathcal{U}_{\mathrm{HP}} \subseteq \mathcal{U}$.
 \item {Standard-Priority (SP) Users:} Representing users with applications such as web browsing, email, or any non-critical data service that generally require a low bandwidth/ data rate. The group of SP users is denoted by $\mathcal{U}_{\mathrm{SP}} \subseteq \mathcal{U}$.
\end{itemize}
We assume these groups form a partition of the total user group $\mathcal{U}$, such that $\mathcal{U}_{\mathrm{HP}} \cap \mathcal{U}_{\mathrm{SP}} = \emptyset$ and $\mathcal{U}_{\mathrm{HP}} \cup \mathcal{U}_{\mathrm{SP}} = \mathcal{U}$. In this work we assumed that users are assigned to a group based on predefined ratios. The specific and measurable QoS requirements are defined for each group as follows:
\begin{itemize}
    \item { HP Users ($u \in \mathcal{U}_{\mathrm{HP}}$):}
        \begin{itemize}
            \item Minimum required data rate: $R_{\mathrm{HP}}^{\mathrm{req}}$ .
        \end{itemize}
    \item { SP Users ($u \in \mathcal{U}_{\mathrm{SP}}$):}
        \begin{itemize}
            \item Minimum required data rate: $R_{\mathrm{SP}}^{\mathrm{req}}$  where ( $R_{\mathrm{SP}}^{\mathrm{req}} < R_{\mathrm{HP}}^{\mathrm{req}}$).
        \end{itemize}
    \item {Service Quality (Outage):}
        \begin{itemize}
           \item  QoS also implies providing a quality level of service. This is directly related to the \textit{Outage Probability}, defined as the likelihood that a SP user's achieved data rate falls below a minimum acceptable threshold ($R_{\mathrm{outage}}$). Minimizing outage is crucial for the quality of service delivered for users in the VLC network.
        \end{itemize}
\end{itemize}

These QoS definitions are fundamental to the MARL framework design. The MARL agents (VLC APs) must learn resource allocation policies (power $P_{a,u}(t)$, user selection $\mathcal{U}_a(t)$) that strive to meet these objectives. The agents' observations include information related to recent QoS performance. Critically, the reward function guiding the agents' learning process will be directly based on meeting these QoS requirements. For instance, agents receive positive rewards for achieving the target data rates ($R_{\mathrm{HP}}^{\mathrm{req}}, R_{\mathrm{SP}}^{\mathrm{req}}$) for their associated users and incur penalties for failing to meet minimum rate requirements. Eventually, the MARL framework will learn to balance these potentially conflicting QoS goals across all users while simultaneously managing network stability objectives (throughput, interference, handovers), a common trade-off in network resource allocations.

\subsection{Network Stability}
\label{subsec:stability_metrics}

Besides satisfying the diverse network QoS needs, maintaining overall network stability is a key objective, particularly under the dynamic conditions resulting from user mobility and high-density deployments. In this work, we consider the following network stability components:

\begin{itemize}
\item \textbf{}Handover Stability: Effective mobility management includes the reduction of disruptions resulting from handovers. Key indicators are \cite{LopezPerez2011HetNetHO}:
        \begin{itemize} \item \textit{Handover Rate (HOR)}: A measure of the frequency of handovers per user. Excessively high handover rates can indicate instability. The goal is to achieve necessary handovers efficiently.
           \item \textit{Ping-Pong Ratio (PPR):} The percentage of handovers promptly followed by an additional handover to the previous VLC AP within a specified time period. High PPR indicates instability, which needs to be avoided.
        \end{itemize}
    \item {Network Efficiency:} The overall network efficiency is measured by the \textit {Network Sum-Rate}, which is considered a crucial metric for network stability and effective resource use throughout the network.
\end{itemize}

\subsection{Dimming Model}
\label{subsec:dimming_model}

VLC APs, light sources, inherently combine illumination and communication functions. Dimming control allows adjustment of brightness levels for user comfort, which inevitably interacts with communication performance \cite{Ghassemlooy2017OWCBook}.

We model dimming control through a dimming factor, $\gamma_a$, for each VLC AP $a$. This factor represents the target brightness level relative to the maximum illumination level. In this work, we assume $\gamma_a$ is selected from a discrete set $\Gamma = \{0.2, 0.3, \dots, 1.0\}$. The $\gamma_{\mathrm{min}}$ is set to 0.2 to achieve the minimum level of illumination. We assume that the average transmitted optical power scales proportionally with $\gamma_a$, determining the illumination level in the environment. It is worth to mention that, the overall illumination is subject to minimum requirements based on standards (office space)\cite{ISO8995}.

The communication signal is transmitted concurrently with the illumination function. While various dimming techniques exist and can interact with the communication signal in complex ways \cite{Stefan2013DimmingVLC}. In this work, we assume that the primary impact of operating at a specific dimming level $\gamma_a$ is implicitly handled within the maximum electrical power budget $P_{a,\mathrm{max}}^{\mathrm{elec}}$. The MARL agents must learn to perform resource allocation effectively within this budget and adapt their strategy based on the available power headroom implied by the dimming level. This allows the agent's policy to be conditioned on the lighting requirement. In our MARL frameworks, the dimming level $\gamma_a$ is considered part of the observable state for each agent $a$. It provides crucial context about the VLC AP's operational mode ($P_{a,\mathrm{max}}^{\mathrm{elec}}$). It is worth mentioning that the MARL agents do not control the dimming level. The agents learn, through interaction with the environment, how the current dimming level affects the relationship between their actions (power allocation $P_{a,u}(t)$, user selection $\mathcal{U}_a(t)$ within the budget $P_{a,\mathrm{max}}^{\mathrm{elec}}$) and the resulting outcomes (QoS metrics and stability indicators).

By incorporating the dimming factor, the MARL agents can learn context-aware resource allocation strategies that optimize communication objectives (QoS, stability) effectively under different, pre-defined illumination conditions.

\section{Proposed MARL Framework}
\label{sec:marl_framework}
\subsection{MARL Problem Formulation}
\label{subsec:marl_formulation}

We formulate the dynamic resource allocation problem in the VLC-NOMA network as an MARL task. The MARL key components are defined as follows:

\subsubsection{Agents}
The learning agents are the VLC APs, indexed by $a \in \mathcal{A} = \{1, 2, ..., N_{\mathrm{AP}}\}$. Each agent $a$ learns its own policy $\pi_a$ to engage with the environment and other agents. The agents aim to optimize their objectives, potentially in a collaborative or independent way.

\subsubsection{State Space (Observation)}
At each discrete time step $t$, each agent $a$ receives a local observation $o_a(t)$, which forms its current state representation $s_a(t)$.
This observation includes necessary information for resource allocation assumed in this work:
\begin{itemize}
    \item {Local Channel Gains:} Locally measure channel power gains $G_{a,u}(t)$ for users $u$ currently associated with VLC AP $a$ ($\mathcal{U}_a(t)$)
    \item {Neighbor Channel Gains:} Locally measure channel power gains $G_{a',u}(t)$ from relevant neighboring VLC APs $a'$ to associated users $u \in \mathcal{U}_a(t)$.
    \item {Interference Information:} Locally measure ICI affecting the associated users.
    \item {QoS Information:} Obtain the current QoS for associated users $u \in \mathcal{U}_a(t)$. Such as users achieved data rate in the previous step (in the environment) $\mathrm{Rate}_{a,u}(t-\Delta t)$ and their QoS class (HP or SP user).
    \item {Dimming Level:} The operational dimming factor $\gamma_a$ of light sources (VLC AP $a$).
    \item {Handover Status:} Handover information for associated users include the time since last handover and previous VLC AP to optimize the handover ping-pong effect.
    \item {Own Previous Action:} The action $a_a(t-\Delta t)$ taken by agent $a$ in the previous step.
\end{itemize}
The state vector $s_a(t)$ integrates this information to provide agent $a$ with context for joint resource allocation and optimization.

\subsubsection{Action Space}
At time step $t$, agent $a$ selects an action $a_a(t) \in A_a$. The action space $A_a$ of VLC-NOMA resource allocations involves:
\begin{itemize}
    \item {User Selection:} Choosing a group of users $\mathcal{U}_a(t) \subseteq \mathcal{U}_{\mathrm{cand},a}(t)$ (candidate users) to serve up to a maximum of $K_{\mathrm{max}}$ users. This is seen as a discrete action.
    \item {Power Allocation:} Assigning NOMA power levels $P_{a,u}(t)$ for each selected user $u \in \mathcal{U}_a(t)$. This is subjected to the total power budget $ P_{a,\mathrm{max}}^{\mathrm{elec}}$ governed by the current dimming factor $\gamma_a$. This is seen as a continuous action.
    \item {Handover Trigger:} For each associated user $u \in \mathcal{U}_a(t)$, an action is decided to whether to trigger a handover procedure. This is seen as a discrete action.
\end{itemize}
The action $a_a(t)$ is thus a composite action including user selection, power allocation, and handover triggers. The dimensionality and nature (discrete/continuous) of the action space components have been carefully considered during algorithm implementation in this paper.

\subsubsection{Reward Function}
The reward $r_a(t)$ received by agent $a$ (or the global reward $r(t)$ used in centralized training) is designed to guide the learning process of resource allocations towards achieving high QoS reliability and network stability. We define a weighted-sum reward structure:
\begin{equation}
\label{eq:reward_unified}
r_a(t) = w_{\mathrm{QoS}} \cdot R_{\mathrm{QoS},a}(t) + w_{\mathrm{Stab}} \cdot R_{\mathrm{Stab},a}(t)
\end{equation}
where $w_{\mathrm{QoS}}$ and $w_{\mathrm{Stab}}$ are primary weighting factors balancing the two main objectives. The components are defined as follows:
\begin{itemize}
    \item {QoS Reward/ Penalty ($R_{\mathrm{QoS},a}(t)$):} Focuses on meeting specific data rate requirements for HP users and ensuring baseline quality via outage avoidance for SP users.
        \begin{itemize}
            \item \textit{HP Rate Satisfaction:} A positive reward for each HP user meeting its target rate, given by $ +w_{\mathrm{HP\_met}} \times (\text{\#HP users satisfaction rate}) $.
            \item \textit{Outage Penalty:} A negative reward applied primarily to users who fall below the outage threshold, given by $ -w_{\mathrm{outage}} \times (\text{\#SP users dissatisfaction rate}) $.
        \end{itemize}

    \item {Stability Reward/Penalty ($R_{\mathrm{Stab},a}(t)$):} Focuses on handover stability and overall network efficiency.
        \begin{itemize}
            \item \textit{Handover Frequency Penalty:} A negative reward penalizing initiated handovers, equal to $ -w_{\mathrm{HO}} \times (\text{\# HOs initiated by agent } a \text{ at step } t) $.
            \item \textit{Ping-Pong Handover Penalty:} A negative reward for handovers identified as ping-pong events (based on HO history in state), equal to $ -w_{\mathrm{pp}} \times (\text{\# Ping-Pong HOs involving agent } a \text{ at step } t) $.
            \item \textit{Sum Throughput Reward:} A positive term encouraging overall network efficiency, given by $ +w_{\mathrm{thr}} \times (\text{scaled achieved throughput}) $.
        \end{itemize}
\end{itemize}
The effective performance of the MARL agents are critically depending on the careful tuning and scaling of the main weights ($w_{\mathrm{QoS}}, w_{\mathrm{Stab}}$) and internal sub-weights ($w_{\mathrm{HP\_met}}, w_{\mathrm{outage}}, w_{\mathrm{HO}}, w_{\mathrm{pp}}, w_{\mathrm{thr}}$). We employed a shaped reward function to offer the MARL agents more granular feedback and improve learning efficacy. This approach scales rewards in proportion to the level of QoS satisfaction attained by users and network stability requirements. Thus creating denser learning signals. The specific design, integration, and tuning of this shaped reward mechanism are detailed in Sections IV-C and V-B

\subsubsection{Objective}
The goal is to find optimal policies (for resource allocations) $\pi^* = \{\pi_1^*, ..., \pi_{N_{\mathrm{AP}}}^*\}$ that maximize the expected long-term discounted cumulative reward for the system\cite{Sutton2018}:
\begin{equation}
\label{eq:objective}
\pi^* = \arg\max_{\pi} \mathbb{E}_{\pi} \left[ \sum_{k=t}^{T} \delta^{k-t} r_k \mid s_t \right]
\end{equation}
where $\mathbb{E}_{\pi}[\cdot]$ denotes the expectation under the joint policy $\pi = \{\pi_1, ..., \pi_{N_{\mathrm{AP}}}\}$, $\delta \in [0, 1)$ is the discount factor controlling the importance of future rewards, $T$ is the time horizon and $r_k$ represents the relevant reward at time step $t$, given the state $s_a(t)$.

\subsection{MARL Frameworks}
\label{subsec:marl_arch_algos}

In this section, we present the selected MARL framework to optimize the resource allocations in VLC-NOMA network. We aim to compare the effectiveness of two well-known MARL architectures: Centralized Training with Centralized Execution (CTCE) and Centralized Training with Decentralized Execution (CTDE). Both architectures are coupled with Proximal Policy Optimization (PPO) algorithm.

\subsubsection{CTDE architecture with Multi-Agent Proximal Policy Optimization (MAPPO)}
\label{ssubsec:ctde_mappo}

The CTDE architecture is suitable for a dynamic VLC-NOMA network. Where VLC APs agents must promptly adapt to the channel fluctuations due to user mobility and light dimming changes through CTDE decentralized execution. The CTDE centralized training enables agents to develop coordinated tactics essential for tackling network-wide issues, including interference and fulfilling diverse QoS requirements \cite{Sadeghi2023MARLComNetSurvey}. We adopt MAPPO \cite{Yu2022MAPPO}, an algorithm recognized for its stability and robust performance in cooperative MARL implementation. MAPPO's architecture enables decentralized agents to acquire policies that implicitly coordinate (directed by a centralized critic), making it effective for optimizing resource allocations in VLC-NOMA networks. In our MAPPO (CTDE) implementation:
\begin{itemize}
    \item Each VLC AP agent $a$ employs a decentralized actor network $\pi_{\theta_a}(a_a|s_a(t))$. This network takes only the local state $s_a(t)$ as input to generate the composite resource allocation and handover action $a_a(t)$. The agents actors do not share parameters, ensuring decentralized execution.
    \item A centralized critic network $V_{\phi}(s_{\mathrm{global}}(t))$ is used during the training phase. The global state $s_{\mathrm{global}}(t)$ provided to the critic is a concatenation of all individual local agent states $\{s_1(t), \dots, s_{N_{\mathrm{AP}}}(t)\}$. This allows the critic to learn an accurate value function reflecting the overall system state and joint agent performance.
    \item During training, the centralized critic provides a stable and comprehensive learning signal for updating the decentralized actors. Each actor $\pi_{\theta_a}$ is updated using the PPO objective function, leveraging advantage estimates calculated from the centralized value function $V_{\phi}$. MAPPO's effectiveness has been demonstrated in complex resource allocation tasks, such as Radio Access Network (RAN) slicing \cite{Hua2021MAPPO_RANslicing}.
    \item During execution, only the trained decentralized actor networks $\pi_{\theta_a}$ are deployed. Each VLC AP makes decisions based solely on its local state $s_a(t)$, preserving the scalability and low-latency benefits essential for real-time operation in dynamic VLC-NOMA networks.
\end{itemize}
It is worth mentioning that the centralized training phase for MAPPO (CTDE) can still present challenges if the dimensionality of the global state $s_{\mathrm{global}}(t)$ for the critic becomes excessively large in very extensive networks.

\subsubsection{CTCE architecture with Centralized PPO (CenPPO)}
\label{ssubsec:ctce_cen_ppo}

The CTCE architecture can be considered as a valuable benchmark for this work. As CTCD decisions can be made based on a comprehensive global network view at execution time. This is particularly relevant in VLC-NOMA networks where VLC APs' actions are tightly coupled through ICI \cite{Sadeghi2023MARLComNetSurvey}. We implement CTCE using a single, centralized PPO agent \cite{Schulman2017PPO} that holistically controls all VLC APs. In our CenPPO (CTCE) implementation:
\begin{itemize}
    \item A single logical agent acts as the central controller for all VLC APs in the VLC-NOMA network. It observes the concatenated global state $s_{\mathrm{global}}(t) = \{s_1(t), \dots, s_{N_{\mathrm{AP}}}(t)\}$, which comprises the local states of all VLC APs.
    \item The VLC AP agent's actor network is $\pi_{\theta}(\mathbf{a}|s_{\mathrm{global}}(t))$ and critic network is $V_{\phi}(s_{\mathrm{global}}(t))$. Both networks receive the full global state $s_{\mathrm{global}}(t)$ as input.
    \item The actor network $\pi_{\theta}$ generates the joint action $\mathbf{a}(t) = \{a_1(t), \dots, a_{N_{\mathrm{AP}}}(t)\}$ for all VLC APs simultaneously. This joint action vector encapsulates the individual composite actions. The central agent explicitly uses this to make decisions that aim to mitigate interference and optimize overall network objectives.
    \item Training employs the standard PPO algorithm, using the global state, the joint action, and a system-level reward which is applied globally.
    \item During execution, the central controller gathers the global state $s_{\mathrm{global}}(t)$, uses the trained policy $\pi_{\theta}$ to compute the joint action $\mathbf{a}(t)$, and then distributes the corresponding individual action $a_a(t)$ to each VLC AP $a$. This allows for explicit coordination based on the full network picture.
\end{itemize}
Centralized controllers using PPO and similar algorithms have been applied to related wireless resource management problems \cite{Li2022CentPPO_UAV}. The primary drawback of CTCE is scalability. The global state and joint action spaces grow significantly with the number of VLC APs ($N_{\mathrm{AP}}$) and users. This can limit its practical applicability in very large-scale VLC networks and impose considerable communication overhead. However, it provides a useful performance upper bound for comparison in our simulated VLC-NOMA scenario.

\subsubsection{Neural Network layout (NN)}
\label{ssubsec:nn_arch}
\begin{figure}[htbp]
    \centering
    \includegraphics[width=1\columnwidth]{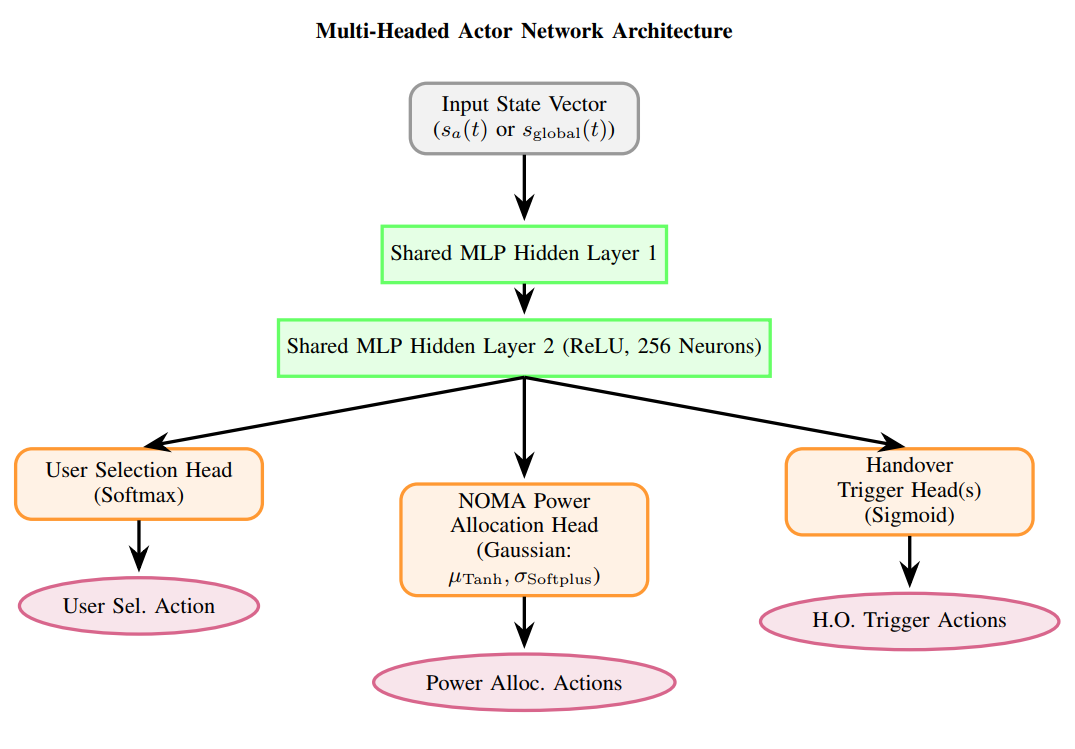}
    \caption{Structure of the Multi-Headed Actor Network.}
    \label{fig:multi_headed_actor_network}
\end{figure}

All actor and critic networks within both MAPPO (CTDE) and CenPPO (CTCE) frameworks are implemented using Multi-Layer Perceptrons (MLPs). The MLPs has 2 hidden layers (each containing 256 neurons with ReLU activation) were used for both actors and critics. Actor outputs employed appropriate activation functions Softmax for discrete selections, Gaussian policy heads for normalized power allocation, and Sigmoid for handover triggers. While critic outputs used linear activation. The final layer(s) of actor networks are structured as separate output heads originating from the shared MLP body:
\begin{itemize}
            \item For \textit{discrete user selection} (choosing up to $K_{\mathrm{max}}$ users from candidates), a head with a softmax activation for selecting each candidate user.
            \item For \textit{continuous NOMA power allocation} (power levels $P_{a,u}(t)$ to selected users), a head models a Gaussian distribution. It has two sub-heads: one outputting the mean values $\mu$ (passed through a Tanh activation and scaled to ensure power constraints and positivity) and another outputting values passed through a softplus activation to ensure positive standard deviations $\sigma$ for each selected user's power.
            \item For \textit{discrete handover trigger} decisions for each associated user ( keep current association with VLC AP or initiate handover), a head with a sigmoid activation function is used, allowing for independent binary trigger decisions.
\end{itemize}
It is worth mentioning that, no parameter sharing is employed between the actor networks of different agents in the MAPPO framework, nor between actors and critics beyond the standard PPO structure. Fig. \ref{fig:multi_headed_actor_network} illustrates detailed of the shared MLP body and distinct output heads for user selection, NOMA power allocation, and handover triggers. Specific NN dimensions and hyperparameters are detailed in Section \ref{subsec:sim_env}.

\begin{figure}[htbp]
    \centering
    \includegraphics[width=1\columnwidth]{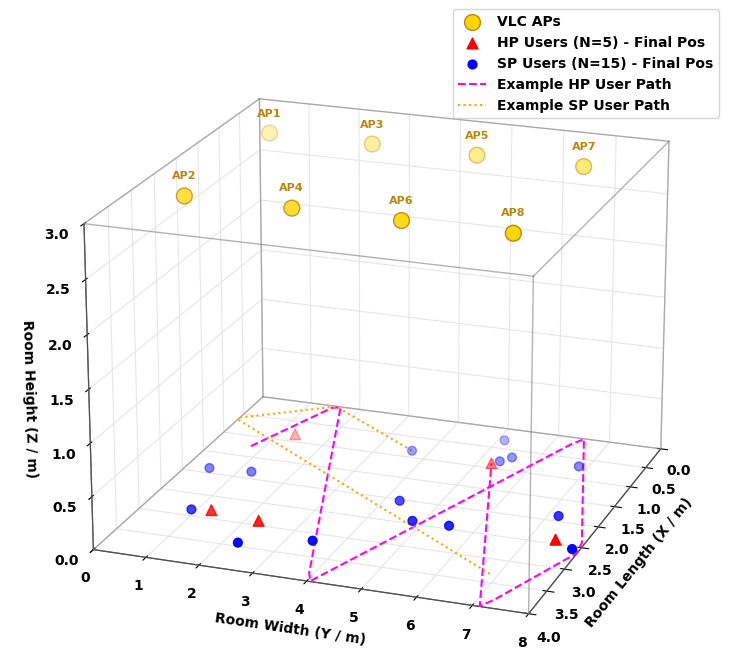}
    \caption{3D System model illustrating the indoor environment ($L \times W \times H$) with ceiling-mounted VLC APs, final user positions (HP/SP differentiated), and example user paths.}
    \label{fig:system_model_3d_paths}
\end{figure}

\section{Simulation Setup}
\label{sec:simulation_setup}

This section details the simulation environment, network parameters, MARL framework implementations, benchmark and baseline methods. Comparisons and performance metrics were employed to evaluate the proposed tailored frameworks MAPPO and CenPPO for optimizing resource allocations in VLC-NOMA network. Figure~\ref{fig:system_model_3d_paths} shows the VLC network model setup.

\subsection{Simulation Environment}
\label{subsec:sim_env}

Simulations were conducted using Python 3. The core MARL frameworks were implemented using the PyTorch deep learning framework \cite{paszke2019pytorch}. Training was accelerated using NVIDIA CUDA-enabled GPUs. The dynamic environment encompassing VLC channel models, NOMA protocol , user mobility, QoS classes, and network stability, were developed as a custom simulation environment, interacting with the PyTorch-based agents via a standard RL interface. The simulation progresses in discrete time steps $\Delta t$. Parameters are selected based on typical values and standards for indoor VLC, NOMA, and mobility modeling in \cite{Ghassemlooy2017OWCBook, Komine2004VLCChannel, Ding2017NOMASurvey, ITU-TY1541,Yu2022MAPPO, Schulman2017PPO}. Key parameters are summarized in Table \ref{tab:sim_params}.

\begin{table}[htbp]
\caption{Simulation Parameters}
\label{tab:sim_params}
\centering
\sisetup{reset-text-series = false, text-series-to-math = true, reset-text-family = false, text-family-to-math = true} 
\begin{tabular}{l S[table-format=3.5] l}
\textbf{Parameter} & \textbf{Value} & \textbf{Unit} \\
\midrule 
\textit{Environment} & & \\
Room Dimensions ($L \times W \times H$) & {4x8x3} & m \\
\addlinespace
\midrule
\textit{VLC APs} & & \\
Number of APs ($N_{\mathrm{AP}}$) & 8 & \\
AP Arrangement & {2x4 grid x(1,3), y(1,3,5,7)} \\
AP Height ($H$) & 3 & m \\
Max Electrical Power ($P_{a,\mathrm{max}}^{\mathrm{elec}}$) & 15 & W \\
LED Semi-Angle ($\Phi_{1/2}$) & 60 & $^\circ$ \\
\addlinespace
\midrule
\textit{Users (UEs)} & & \\
Number of Users ($N_{\mathrm{UE}}$) & 20 & \\
UE Height ($H_{\mathrm{UE}}$) & 0.85 & m \\
PD Area ($A_{\mathrm{PD}}$) & \SI{1}{cm^2} & \\
PD FOV ($\Psi_c$) & 70 & $^\circ$ \\
Optical Filter Gain ($T_s$) & 1 & \\
Optical Concentrator Gain ($g$) & 1 & \\
PD Responsivity ($R$) & 0.5 & A/W \\
\addlinespace 
\midrule 
\textit{Channel \& Noise} & & \\

Noise Power Spectral Density ($N_0$) & \num{1e-22} & A$^2$/Hz \\
Modulation Bandwidth ($B$) & 20 & MHz \\

\addlinespace 
\midrule 
\textit{NOMA} & & \\
Max Users per AP ($K_{\mathrm{max}}$) & 10 & \\
SIC Assumption & {Perfect} & \\
\addlinespace 
\midrule 
\textit{Mobility (Random Direction)} & & \\
Min Speed ($v_{\mathrm{min}}$) & 0.5 & m/s \\
Max Speed ($v_{\mathrm{max}}$) & 1.5 & m/s \\
Time Step ($\Delta t$) & 1 & s \\
\addlinespace 
\midrule 
\textit{QoS} & & \\
Percentage HP Users & 25 & \% \\
HP Rate Req ($R_{\mathrm{HP}}^{\mathrm{req}}$) & 12 & Mbps \\
SP Rate Req ($R_{\mathrm{SP}}^{\mathrm{req}}$) & 2 & Mbps \\
Outage Threshold ($R_{\mathrm{outage}}$) & 0.5 & Mbps \\
\addlinespace 
\midrule 
\textit{Dimming} & & \\
Minimum Dimming Factor $\gamma$ & 0.2 & \\
Dimming Factor Set ($\Gamma$) & {\{0.2, ..., 1.0\}} & \\
\addlinespace 
\midrule 
\textit{Handover} & & \\
Ping-Pong Time Window ($T_{\mathrm{pp}}$) & 2 & s \\
Baseline HO Hysteresis & \SI{3}{dB} & \\
\midrule
\textit{MARL Training} & & \\
Agent Type & {CenPPO, MAPPO} & \\
Optimizer & {Adam}\\
Learning Rate & \num{5e-4}\\
Discount Factor ($\gamma$) & 0.97 & \\ 
GAE Lambda ($\lambda_{\mathrm{GAE}}$) & 0.95 & \\
Actor LR & \num{5e-4} & \\
Critic LR & \num{1e-4} & \\
PPO Clip Ratio ($\epsilon_{\mathrm{clip}}$) & 0.2 & \\
Entropy Coefficient & 0.01 & \\
Rollout Buffer Size & 2048 & steps \\
NN Hidden Layer Units & 2 & \\
NN layer Neurons & 256 & \\
Total Training Steps & \num{1.5e6} & steps \\
Evaluation Episodes & 1000 & \\

Max Steps per Episode & 100 & \\
\bottomrule 
\end{tabular}
\end{table}

\subsection{Reward Weight Optimization}
\label{subsec:reward_tuning_bo}

A key aspect of configuring the proposed MARL frameworks is to balance the objectives of QoS and network stability requirements. This primarily controlled by the main weights ($w_{\mathrm{QoS}}$ and $w_{\mathrm{Stab}}$) in the reward function in Eq.~\eqref{eq:reward_unified}. We utilized BO to systematically search for weights that maximize overall system performance governed by a composite score $f(\mathbf{w})$. The reward score obtained from a full train-evaluate cycle which was calculated using the following formula:
\begin{equation}
\label{eq:composite_score}
\begin{split}
f(\mathbf{w}) =w_{\mathrm{QoS}}(& 1.0 \cdot QoSSR_{\mathrm{HP}} - 0.5 \cdot P_{\mathrm{out}}) \\
                 &+ w_{\mathrm{Stab}}( - 0.2 \cdot {HOR}
                 - 0.3 \cdot {PPR} \\
                 & + 0.1 \cdot \bar{R}_{\mathrm{sum}})
\end{split}
\end{equation}
$\bar{R}_{\mathrm{sum}}$ is the normalized value of network rate used in the calculation and scaling of the composite reward score given as $\bar{R}_{\mathrm{sum}} = {R_{\mathrm{sum}}}/{R_{\mathrm{max}}}$
where $R_{\mathrm{max}}$ is the maximum network sum rate used for scaling. By scaling $R_{\mathrm{sum}}$ to a consistent range (0-1) its contribution to the reward score becomes independent of the raw Mbps values achieved. Table \ref{tab:score_term_priority} coefficients are set to enable the VLC-NOMA network to prioritize reliable high data rate services. Such as real-time video conferencing service for HP users. The weights chosen to emphasize HP QoS ($QoSSR_{\mathrm{HP}}$) as critical (1.0), followed by network reliability of ($P_{\mathrm{out}}$) (0.5). Ping-pong ($PPR$, 0.3), handover stability ($HOR$, 0.2), and total throughput ($R_{\mathrm{sum}}$, 0.1) are deemed secondary to ensure primary application performance.

\begin{table}[htbp]
    \centering
    \caption{and Sub-weights coefficients, Reward Terms and Priority Levels}
    \label{tab:score_term_priority}
    \begin{tabular}{@{} l l c @{}}
        \toprule
        \textbf{Sub-weight} & \textbf{Reward Term} & \textbf{Priority Level} \\
        \midrule
        $w_{\mathrm{HP\_met}}=1$    & $QoSSR_{\mathrm{HP}}$      & Highest  \\
        $w_{\mathrm{outage}}=0.5$ & $P_{\mathrm{out, SP}}$  & High     \\
        $w_{PPR}=0.3$               & $PPR$                     & Moderate \\
        $w_{\mathrm{HO}}=0.2$               & $HOR$     & Lower    \\
        $w_{\mathrm{thr}}=0.1$    & $\bar R_{\mathrm{sum}}$       & Lowest   \\
        \bottomrule
    \end{tabular}
\end{table}

\begin{figure}[htbp]
	 \centering	 \includegraphics[width=1\linewidth]{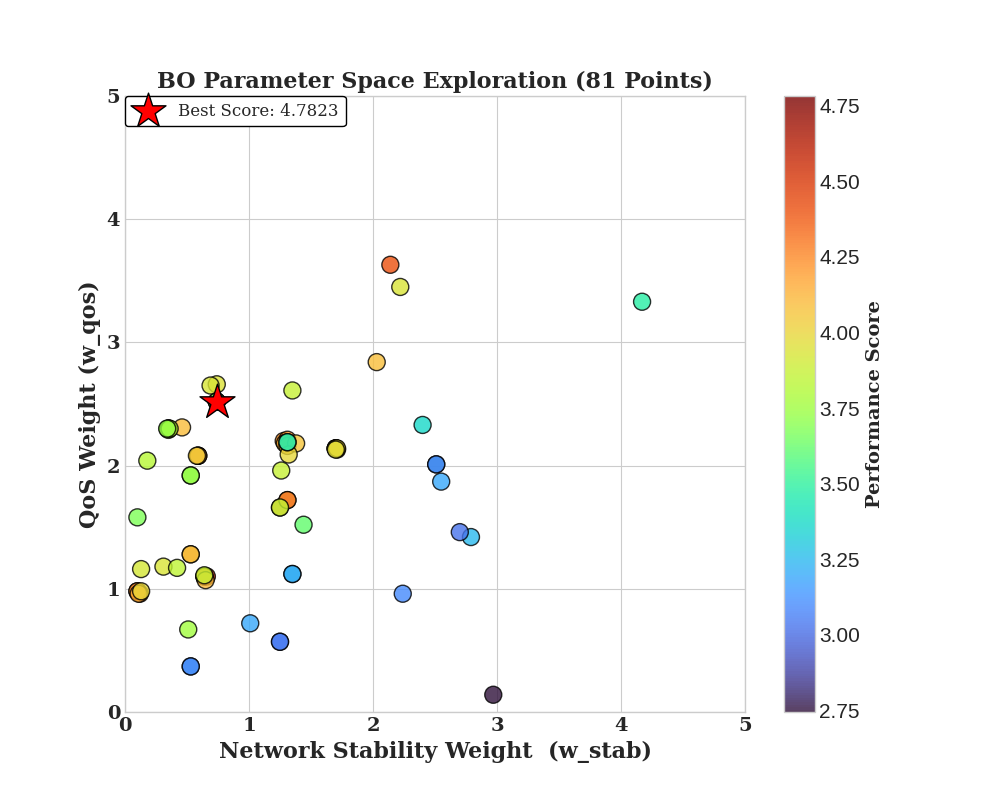}
	 \caption{BO exploration of the reward main weight parameter space ($w_{\mathrm{Stab}}$ vs $w_{\mathrm{QoS}}$). }
	 \label{fig:bo_plot}
\end{figure}

The BO optimization explored the search space $\mathcal{W}$ for $w_{\mathrm{QoS}} \in [0.1, 5.0]$ and $w_{\mathrm{Stab}} \in [0.1, 5.0]$, evaluating 81 points/runs in total. The exploration of the parameter space and the corresponding performance scores are presented in Figure~\ref{fig:bo_plot}.

The BO results indicate that the composite performance score is sensitive to the choice of weights. As shown in Figure~\ref{fig:bo_plot}, the highest performance scores, Target $\approx \num{4.6}-\num{4.8}$, were achieved in specific regions. The optimization converged to a maximum observed performance score of approximately. \num{4.75} at weights $w_{\mathrm{QoS}} \approx \num{2.5}$ and $w_{\mathrm{Stab}} \approx \num{0.75}$. The selection of the BO-tuned weights reflects an emphasis placed on QoS metrics relative to network stability based on the composite score in Eq.~\eqref{eq:composite_score}.

\begin{algorithm}[htbp]
\caption{Bayesian Optimization for Main Reward Weight Tuning}
\label{alg:bo_tuning}
\begin{algorithmic}[1]
\Require Search space $\mathcal{W}$ for $\mathbf{w}=[w_{\mathrm{QoS}}, w_{\mathrm{Stab}}]$
\Require Objective function $f(\mathbf{w})$ (executes MARL train+eval, returns composite score)
\Require Number of initial samples $N_{\mathrm{init}}$
\Require Total number of BO iterations $N_{\mathrm{trials}}$
\Require Acquisition function $\alpha(\mathbf{w})$
\Ensure Optimized main weights $\mathbf{w}^*$

\State \textbf{Initialization:} Sample $N_{\mathrm{init}}$ points $\{\mathbf{w}_1, ..., \mathbf{w}_{N_{\mathrm{init}}}\}$ from $\mathcal{W}$
Evaluate $y_i = f(\mathbf{w}_i)$ for $i = 1, ..., N_{\mathrm{init}}$. Initialize dataset $D = \{(\mathbf{w}_i, y_i)\}_{i=1}^{N_{\mathrm{init}}}$.
\State \textbf{Optimization Loop:}
\For{$t = N_{\mathrm{init}}$ to $N_{\mathrm{trials}}-1$}
    \State Fit/Update Gaussian Process model using data $D$.
    \State Find $\mathbf{w}_{t+1} = \arg\max_{\mathbf{w} \in \mathcal{W}} \alpha(\mathbf{w} | D)$.
    \State Evaluate $y_{t+1} = f(\mathbf{w}_{t+1})$. \Comment{Requires full MARL train+eval}
    \State Augment dataset $D \leftarrow D \cup \{(\mathbf{w}_{t+1}, y_{t+1})\}$.
\EndFor
\State \textbf{Return Best:} Find $\mathbf{w}^* = \arg\max_{\mathbf{w}_i | (\mathbf{w}_i, y_i) \in D} y_i$.
\end{algorithmic}
\end{algorithm}

\subsection{Training Performance (MAPPO vs CenPPO)}

\begin{figure}[htbp]
	 \centering
		 \includegraphics[width=1\linewidth]{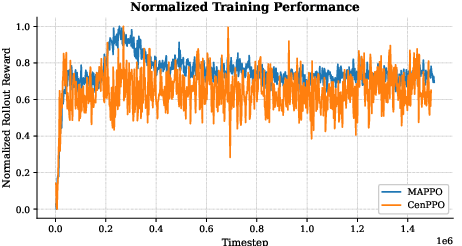}
	 \caption{Normalized Training Performance showing Reward versus Timestep for MAPPO and CenPPO.}
	 \label{fig:training_plot}
\end{figure}

Figure~\ref{fig:training_plot} shows the learning process of the proposed MAPPO (CTDE) and CenPPO (CTCE) frameworks. The figure plots the training normalized cumulative reward against the total environment steps for MAPPO and CenPPO. Both MARL frameworks show converging towards stable reward levels, learning trends indicating successful policy optimization concerning the reward function. MAPPO appears to converge slightly faster as compared to CenPPO.

\subsection{Comparative Performance Analysis (MAPPO vs CenPPO vs Baseline)}

\begin{figure}[htbp]
	 \centering
	 \includegraphics[width=1\linewidth]{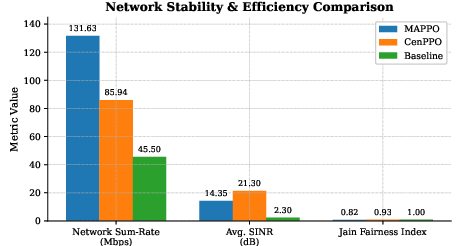}
	 \caption{Comparison of Network Stability Metrics: Network Sum-Rate, Average SINR, and JFI.}
	 \label{fig:stability_plot}
\end{figure}

\begin{figure}[htbp]
    \centering
    \includegraphics[width=9cm, height=7cm]{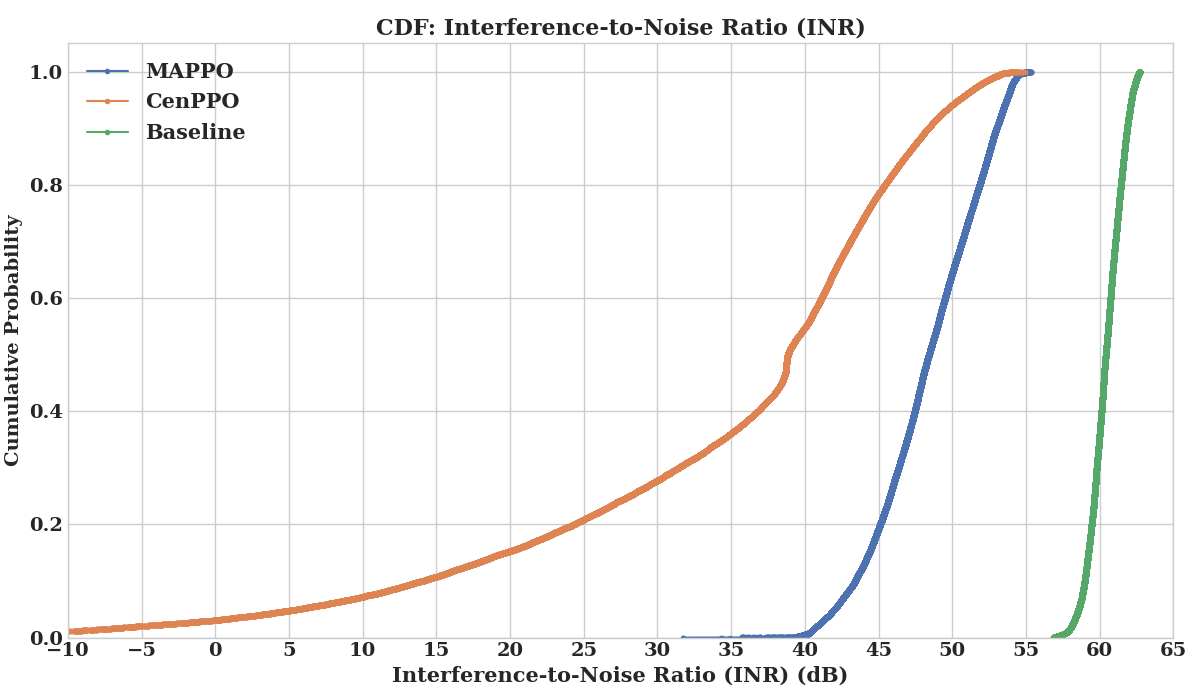}
    \caption{CDF of User Interference-to-Noise Ratio (INR) for MAPPO, CenPPO, and Baseline.}
    \label{fig:inr_cdf_plot}
\end{figure}

The core performance comparison between the proposed MAPPO (CTDE) and CenPPO (CTCE) and the Baseline resources allocation approaches is presented in Figures~\ref{fig:qos_plot} to~\ref{fig:power_plot}. The analysis focuses on assessing how effectively each resources allocation approach achieves the joint objectives of QoS satisfaction and network stability under the challenging dynamic conditions.

\subsubsection{QoS Performance Comparison}

\begin{figure}[htbp]
	 \centering
	 \includegraphics[width=1\linewidth]{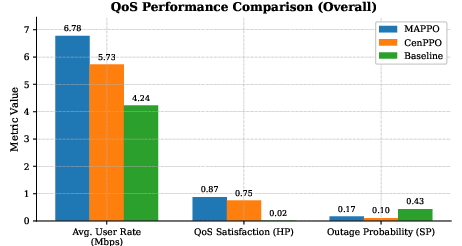}
	 \caption{Comparison of Network QoS metrics: Average User Rate, QoSSR HP, and SP Outage Probability.}
	 \label{fig:qos_plot}
\end{figure}

\begin{figure}[htbp]
	 \centering
	 \includegraphics[width=1\linewidth]{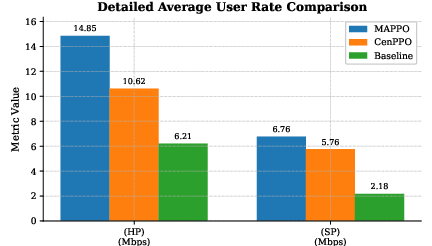}
	 \caption{Comparison of HP and SP users average data rates ($\bar{R}_{\mathrm{HP}}$,$\bar{R}_{\mathrm{SP}}$).}
	 \label{fig:detailed_rate_plot}
\end{figure}

\begin{itemize}
	 \item \textbf{Average User Data Rate:} Figure~\ref{fig:qos_plot} shows that both MAPPO and CenPPO significantly outperform the Baseline approach. MAPPO achieves the highest overall rate of (\SI{6.78}{\mega\bit\per\second}). Followed by CenPPO with (\SI{5.73}{\mega\bit\per\second}). while Baseline archived (\SI{4.24}{\mega\bit\per\second}). Figure~\ref{fig:detailed_rate_plot} shows MAPPO delivers considerably higher rates specifically to HP users (\SI{14.85}{\mega\bit\per\second}) compared to CenPPO (\SI{10.62}{\mega\bit\per\second}) and the Baseline (\SI{6.21}{\mega\bit\per\second}). For SP users, MAPPO (\SI{6.76}{\mega\bit\per\second}) and CenPPO (\SI{5.76}{\mega\bit\per\second}) perform similarly and better than the Baseline with (\SI{2.18}{\mega\bit\per\second}).
	 \item \textbf{QoSSR (HP Users):} Figure~\ref{fig:qos_plot} shows the both MAPPO and CenPPO achieve high satisfaction ratios for HP users which are significantly better than the Baseline. where MAPPO leads with \num{0.87}, closely followed by CenPPO at \num{0.75}. while Basline at (\num{0.02}). These results indicate that the MAPPO is more effective at prioritizing HP users to meet their target.
	 \item \textbf{Outage Probability (SP Users):} Figure~\ref{fig:qos_plot} shows the outage probability for SP users. The CenPPO achieves the lowest SP outage probability of (\num{0.1}. The MAPPO performs reasonably well (\num{0.17}), while the Baseline exhibits the highest outage (\num{0.43}). These results indicate that the CenPPO appears better at ensuring baseline service for the network users.
\end{itemize}
The QoS results (as shown in Figures~\ref{fig:qos_plot},~\ref{fig:detailed_rate_plot}) indicate distinct strategies learned by the MARL agents. The MAPPO tend to prioritize the HP users, potentially leveraging its decentralized nature for faster local optimization, which is leading to higher HP rates and satisfaction. In contrast, the CenPPO achieves a more balanced QoS outcome, sacrificing some HP performance for significantly minimizing the network users' outage. This is due to that CenPPO is benefiting from its global view for resource allocation.

\subsubsection{Network Stability Comparison}

\begin{itemize}
	 \item \textbf{Network Sum-Rate:} Figure~\ref{fig:stability_plot} shows the achieved network sum rates. The MAPPO achieves the highest sum-rate (\SI{131.63}{\mega\bit\per\second}). Followed by CenPPO (\SI{85.94}{\mega\bit\per\second}) and the Baseline (\SI{45.5}{\mega\bit\per\second}). The results indicates that the MAPPO tend to maximize the overall network throughput.
	 \item \textbf{Average SINR:} Figure~\ref{fig:stability_plot} shows the user average SNIR in dB. The CenPPO attains the highest average SINR (\SI{21.3}{\deci\bel}) exceeding the MAPPO (\SI{14.35}{\deci\bel}) and the Baseline (\SI{2.30}{\deci\bel}). These indicate that the CenPPO centralized execution leads to better interference coordination in the network.
	 \item \textbf{Interference-to-Noise Ratio (INR):} Figure~\ref{fig:inr_cdf_plot} shows the CDF of user INR. The CenPPO maintains lower INR levels for most users compared to the MAPPO and the Baseline. CDF at 50 \%, the users under CenPPO framework experience INR below \SI{40}{dB}, whereas for the MAPPO \SI{48}{dB} and Baseline \SI{61}{dB}. These results corroborates the SINR results showing the lead for CenPPO.
	 \item \textbf{Jain's Fairness Index (JFI)} \cite{Jain1984Fairness}: Measures the fairness of power allocation among all users based on their averaged allocated power $\bar{P}_u$. It is calculated as $JFI = \frac{\left(\sum_{u \in \mathcal{U}} \bar{P}_u\right)^2}{N_{\mathrm{UE}} \cdot \sum_{u \in \mathcal{U}} \bar{P}_u^2}$.
\end{itemize}
The CenPPO shows a superior SINR and significantly lower INR performance. This reflect the advantage of the CenPPO centralized execution in managing inter-cell interference across the VLC network. The differing fairness indices highlight the inherent trade-off faced by the agents where the MAPPO's lower fairness is due to its strong prioritization of HP users. While the CenPPO's high fairness indicates a more equitable, globally considered resource allocation strategy.
\subsubsection{Handover Performance Comparison}

\begin{figure}[htbp]
	 \centering
	 \includegraphics[width=0.9\linewidth]{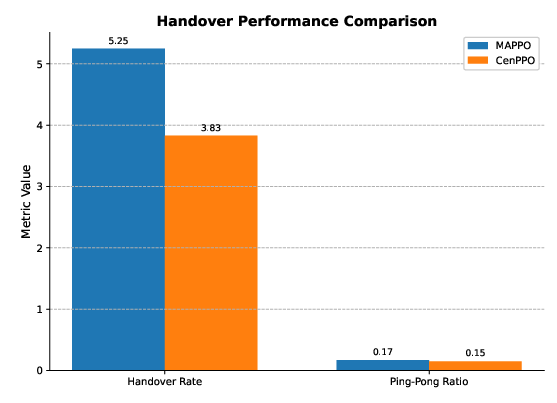}
	 \caption{Comparison of handover performance metrics: Handover Rate (HOR) per user per second and Ping-Pong Ratio (PPR).}
	 \label{fig:handover_plot}
\end{figure}

\begin{itemize}
	 \item \textbf{Handover Rate (HOR):} Figure~\ref{fig:handover_plot} shows the handover performance of the VLC network. The CenPPO achieves a lower HOR of (\num{3.83} HOs/user/s) compared to the MAPPO with (\num{5.25} HOs/user/s).
	 \item \textbf{Ping-Pong Ratio (PPR):} Figure~\ref{fig:handover_plot} shows the handover ping-pong effect performance of the network. The CenPPO shows a handover stability with a lower PPR of (\num{0.15}) compared to MAPPO (\num{0.17}). These results indicate that the CenPPO makes robust decisions by reducing the wasteful back-and-forth user switching between the VLC APs.
\end{itemize}
The CenPPO a high performance in maintaining handover stability, evidenced by both lower HOR and PPR rates. This is due to that the CenPPO global perspective during execution enables more robust handover decisions considering network conditions.
\subsubsection{Power Consumption Comparison}

\begin{figure}[htbp]
	 \centering
	 \includegraphics[width=0.9\linewidth]{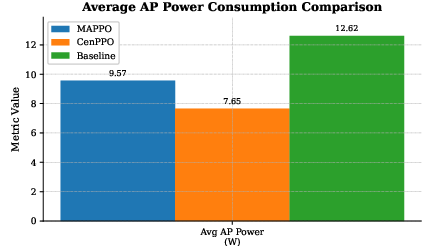}
	 \caption{Comparison of average VLC AP electrical power consumption.}
	 \label{fig:power_plot}
\end{figure}
Figure~\ref{fig:power_plot} shows the average electrical power consumed per VLC AP. The Baseline uses an average of (\SI{12.62}{\watt}). The MAPPO uses less (\SI{9.57}{\watt}). While the CenPPO is the most power-efficient with (\SI{7.65}{\watt}). The lower power consumption of the CenPPO indicates it's excellence in interference coordination and performing resource allocation with less overall transmission power compared to the less coordinated MAPPO and Baseline approaches.

\subsubsection{Computational Complexity Analysis}
Table~\ref{tab:complexity} summarizes the complexity orders for the two proposed MARL frameworks. The MAPPO (CTDE) benefits from decentralized execution and offering lower runtime complexity that scales well with $N_{AP}$. However, the MAPPO centralized training is more demanding. In contrast, the CenPPO (CTCE) shows high complexity in both training and execution due to its reliance on global state processing. This limit the CenPPO scalability.

\begin{table*}[htbp]
\centering
\caption{Computational Complexity Comparison}
\label{tab:complexity}
\begin{tabular}{|l|l|l|l|}
\hline
\textbf{MARL} & \textbf{Phase} & \textbf{Complexity Order} & \textbf{Level} \\
\hline
MAPPO & Execution & $\mathcal{O}(N_{AP} \times f_{NN}(D_s))$ & Low/Moderate \\ \cline{2-4}
& Training & $\mathcal{O}(N_{AP} \times f_{NN}(D_s) + f_{NN}(D_{s,global}))$ & High \\
& & & \\
\hline
CenPPO & Execution & $\mathcal{O}(f_{NN}(D_{s,global}))$ & High \\ \cline{2-4}
& Training & $\mathcal{O}(f_{NN}(D_{s,global}))$ & Very High \\
& & & \\
\hline
\multicolumn{4}{l}{\footnotesize Note: $N_{AP}$: Number of agents; $D_s$: Single agent state dim.; $D_{s,global}$: Global state dim.; $f_{NN}(\cdot)$: NN processing complexity.} \\
\end{tabular}
\end{table*}

\subsection{Discussion and Interpretation}
\label{subsec:discussion}
\begin{itemize}
	 \item \textbf{Synthesis of Findings:} The results show that both MARL frameworks significantly outperform the non-adaptive Baseline approach. The MAPPO (CTDE) emerges as the leader in achieving the highest sum-rate and HP user rates. The CenPPO (CTCE) excels in achieving higher SINR, lower interference, better outage performance, better handover stability and lower power consumption. The Baseline is generally non-competitive due to it basic policy in resource allocations. Except for achieving high fairness due to its simple equal allocation, as assumed.

	 \item \textbf{MAPPO vs. CenPPO Trade-offs:} The MAPPO decentralized execution allows it to react faster locally and achieve higher throughput for prioritized users. However, the MAPPO suffers from higher interference and less stable handovers due to the lack of real-time global coordination during execution. The CenPPO leverages its global view for better interference management, handover stability, fairness, and power efficiency. But at the cost of lower throughput and potential scalability issues in larger dense VLC networks.

	 \item \textbf{Limitations:} In this paper, the results and findings are specific to the simulated environment, parameter settings, mobility model, and the chosen composite score for BO. The MARL frameworks performance could vary with different configurations. It is worth mentioning that the VLC simulation environment was included simplifications: 1) perfect SIC is assumed and 2) a perfect global state assumption for the CenPPO during execution.
\end{itemize}

\section{Conclusion}
\label{sec:conclusion}

This paper tackled a complex challenge of optimizing resource allocation while provisioning diverse Quality of Service (QoS) and network stability in a dynamic indoor Visible Light Communication-Non-Orthogonal Multiple Access (VLC-NOMA) network. To address this complex problem, we proposed, developed, and comparatively evaluated tailored Multi-Agent Reinforcement Learning (MARL) frameworks, specifically investigating CTDE via MAPPO and CTCE via Centralized PPO. Our approach involved designing customized MARL components (state, action, reward) suited for the unique VLC-NOMA dynamics and employing Bayesian Optimization to systematically determine the critical balance between competing QoS and stability objectives within the multi-objective shaped reward function. We adopted a VLC system model incorporating realistic channel characteristics, NOMA principles, user mobility, dynamic interference, and dimming considerations.
The comparison results between MAPPO and CenPPO highlighted the inherent trade-offs:
\begin{itemize}
    \item {MAPPO (CTDE)} excelled in throughput-related metrics achieving approximately {53\% higher network sum-rate}, {40\% higher average rates for HP users}, and {17\% higher rates for SP users} as compared to CenPPO. Also, the MAPPO provided {16\% higher QoS satisfaction for HP users}. However, the MAPPO results in a higher interference environment and exhibit less network stability reflected in a {37\% higher handover rate} and a {14\% higher ping-pong ratio} as compared to CenPPO.
    \item {CenPPO (CTCE)} shows superior performance in network stability. The CenPPO achieved about 7 dB higher average SINR, which indicates better interference management. Also, the CenPPO has {37\% lower outage probability for SP users}, better handover stability with {14\% lower ping-pong ratio}, {37\% lower handover rate} and operated more efficiently with {20\% lower average AP power consumption} as compared to the MAPPO. However, the CenPPO faces inherent scalability limitations in 1)training and execution and high implementation complexity due to its centralized nature.
\end{itemize}

This work underscores the potential of MARL approaches to effectively optimize resources in complex VLC-NOMA networks. The choice between architectures depends on the specific network scale, QoS, stability requirements and the scalability trade-offs of each approach.

\bibliographystyle{IEEEtran} 

\end{document}